\documentclass{iau} 
\usepackage{graphicx}
\usepackage{wrapfig}

\title[The evolution of Sakurai's object] 
{The very fast evolution of Sakurai's object}

\author[G.C. Van de Steene, P. A. M. van Hoof, S. Kimeswenger, et al.]  
{G. C. Van de Steene$^1$
\and P. A. M. van Hoof$^1$
\and S. Kimeswenger$^2$
\and A. A. Zijlstra$^3$
\and A. Avison$^3$
\and L. Guzman-Ramirez$^4$
\and M. Hajduk$^5$
\and F. Herwig$^6$
}

\affiliation{
$^1$Royal Observatory of Belgium, Astronomy \& Astrophysics, Ringlaan 3, Brussels, Belgium \\
email: {\tt g.vandesteene@oma.be} \\ [\affilskip]
$^2$Universidad Católica del Norte, Antofagasta, Chile \\ [\affilskip]
$^3$Jodrell Bank Centre for Astrophysics, Manchester, UK \\ [\affilskip]
$^4$Sterrewacht, Universiteit Leiden, The Netherlands \\ [\affilskip]
$^5$Nicolaus Copernicus Astronomical Center, Torun, Poland \\ [\affilskip]
$^6$University of Victoria, Victoria, Canada \\ [\affilskip]
}

\pubyear{2016}
\volume{323}  
\jname{Planetary nebulae: Multiwavelength probes of stellar and galactic evolution}
\editors{X. Liu, L. Stanghellini, \& A. Karakas eds.}
\begin{document}

\maketitle

\begin{abstract}
V4334 Sgr (a.k.a.\ Sakurai's object) is the central star of an old
planetary nebula that underwent a very late thermal pulse a few years
before its discovery in 1996. We have been monitoring the evolution of
the optical emission line spectrum since 2001. The goal is to improve
the evolutionary models by constraining them with the temporal
evolution of the central star temperature.  In addition the high
resolution spectral observations obtained by X-shooter and ALMA show
the temporal evolution of the different morphological components.
\end{abstract}

\firstsection
\section{Introduction}
V4334 Sgr (a.k.a. Sakurai's object) is the central star of an old
planetary nebula (PN) that underwent a very late thermal pulse (VLTP)
a few years before its discovery in 1996 (\cite[Nakano et
  al.\ 1996]{Nakano96}). During the VLTP it ingested its remaining
hydrogen rich envelope into the helium burning shell and ejected the
processed material shortly afterwards to form a new, hydrogen
deficient nebula inside the old PN. The star brightened consideraby to
become a very cool (born-again) AGB star with a spectrum resembling a
carbon star. After a few years, dust formation started in the new
ejecta and the central star became highly obscured, similar to R~CrB
stars. Emission lines were discovered: first He\,{\sc i} 1083~nm in
1998 (\cite[Eyres et al. 1999]{Eyres99}), later in 2001 also optical
forbidden lines from neutral and singly ionized nitrogen, oxygen and
sulfur (\cite[Kerber et al. 2002]{Kerber02}). Sakurai baffled the
scientific community with its very fast evolution, much faster than
pre-discovery models predicted. The current models (\cite[Lawlor \&
  MacDonald 2003]{Lawlor02}, \cite[Herwig et al. 2011]{Herwig11},
\cite[Herwig et al. 2014]{Herwig14}, \cite[Miller Bertolami et
  al. 2006]{MillerBertolami06}) can be improved by constraining them
with the temporal evolution of the central star temperature.

\section{Observations and results} 
\subsection{Optical monitoring}
\begin{wrapfigure}[14]{l}{0.5\textwidth}
\begin{center}
\includegraphics[width=0.9\linewidth]{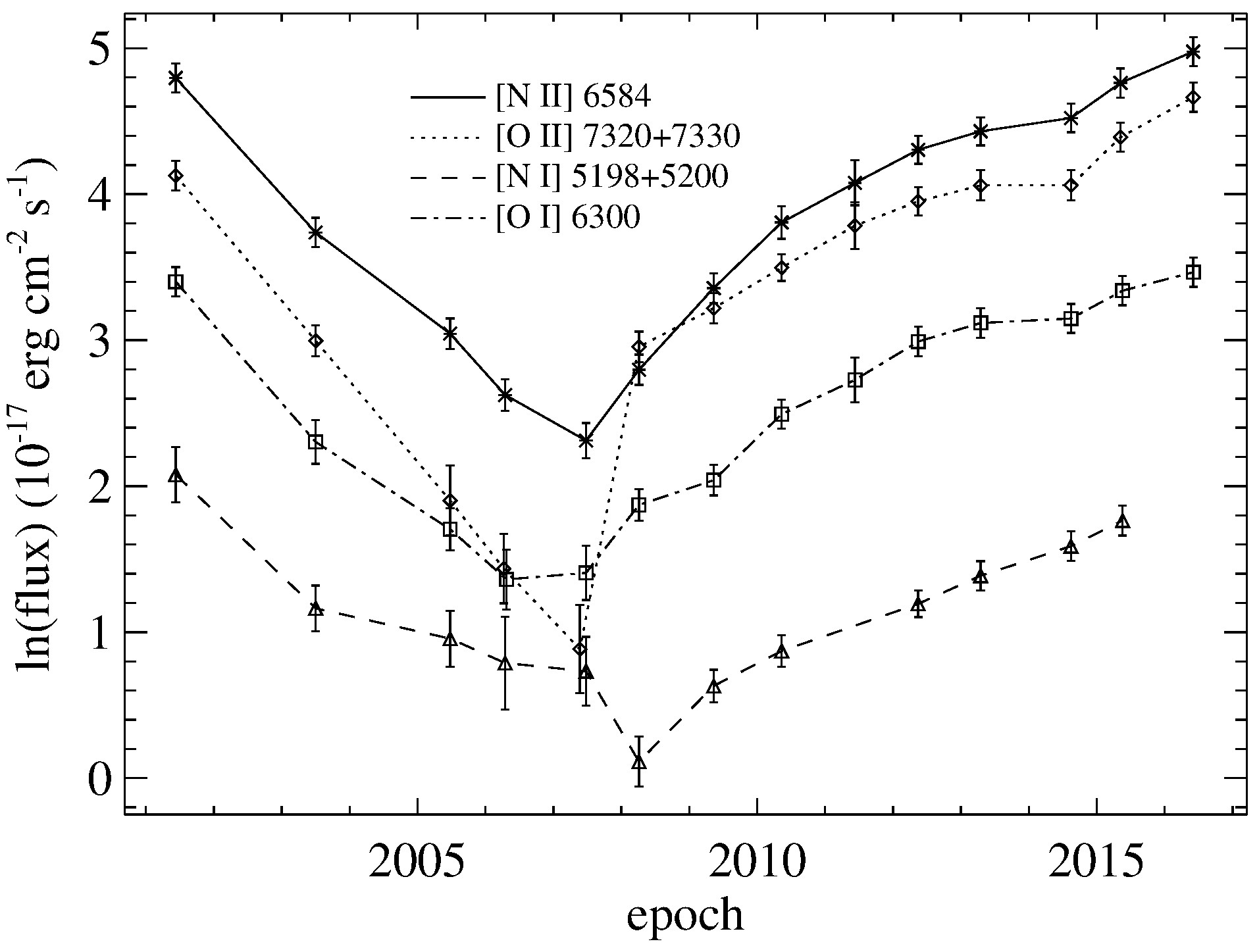} 
\caption{Evolution of emission line fluxes since 2001.}
\label{fig1}
\end{center}
\end{wrapfigure}

We have been monitoring the evolution of the optical emission line
spectrum since 2001 using spectra obtained with FORS at the
ESO-VLT (Fig. \ref{fig1}). 
The goal of this monitoring program is to derive the stellar
temperature as a function of time (\cite[van Hoof et al. 2007]{vHoof07},
\cite[van Hoof et al. 2008]{vHoof08}). From 2001
through 2007 the optical spectrum showed an exponential decline in
flux and the level of excitation also dropped.  We see this as
evidence for a shock that occurred around 1998 and cooled afterwards. A plausible
explanation is that this is the fastest material ejected in the VLTP
hitting older ejecta. The optical line fluxes started to increase
again since 2008. The sudden jump in [O\,{\sc ii}] flux in 2008 could point
to a second shock as the cause of the change in behavior.  The optical
spectrum shows new lines which have been emerging since 2013. Some
have tentatively been identified as electronic transitions of CN. At
least some of these newly emerging lines
are formed very close to the central star, possibly in the disk. In
the PV diagram of [N\,{\sc ii}] 658.3 nm, obtained with X-shooter, we see that the
red- and blue-shifted emission come from different regions that are spatially shifted
w.r.t.\ the continuum. This emission line, as well as other optical forbidden lines, originate in the
bipolar lobes seen by \cite[Hinkle \& Joyce (2014)]{HJ14}.

\subsection{ALMA Observations}

\begin{wrapfigure}[10]{r}{0.40\textwidth}
\vspace*{-0.4cm}
\begin{center}
\includegraphics[width=0.9\linewidth]{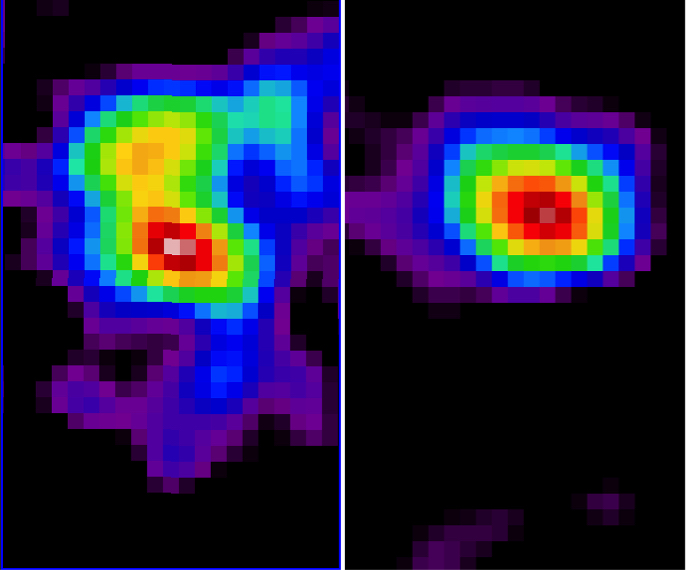} 
\caption{CN (left) and CO emission (right) detected by ALMA in 2015.}
\label{fig2}
\end{center}
\end{wrapfigure}
The continuum emission is unresolved which indicates that all the dust
is in the disk. In the ALMA spectra we detect CO, CN, HC$_3$N, as well as
$^{13}$C isotopologues of these species. The CO and HC$_3$N (+isotopologues) emission
is unresolved, coincides with the position of the central star, and most likely comes
from the disk (Fig.~\ref{fig2}).  The CN and $^{13}$CN emission is spatially resolved and
coincides with the bipolar lobes seen by \cite[Hinkle \& Joyce (2014)]{HJ14} 
(Fig.~\ref{fig2}). CN could be formed via shock-induced dissociation of HCN in the lobes.


\begin{thebibliography}{}

\bibitem[Chesneau et al. (2009)]{Chesneau09}
Chesneau, O., Clayton, G. C., Lykou, F., et al., 2009, \textit{A\&A}, 493, L17
\bibitem[Eyres et al. (1999)]{Eyres99}
Eyres, S. P. S., Smalley, B., Geballe, T. R., et al., 1999, \textit{MNRAS}, 307, L11
\bibitem[Hadjuk et al. (2005)]{Hadjuk05}
Hajduk, M., Zijlstra, A. A., Herwig, F., et al., 2005, \textit{Science}, 308, 231
\bibitem[Herwig et al.  (2011)]{Herwig11}
Herwig, F., Pignatari, M., Woodward, P. R., et al., 2011, \textit{ApJ}, 727, 89
\bibitem[Herwig (2014)]{Herwig14}
Herwig, F., Woodward, P. R., Lin, P.-H., Knox, M., Fryer, C., 2014, \textit{ApJ}, 792, L3
\bibitem[Hinkle \& Joyce (2014)]{HJ14}
Hinkle, K. H., \& Joyce, R. R., 2014, \textit{ApJ}, 785, 146
\bibitem[Kerber et al. (2002)]{Kerber02}
Kerber, F., Pirzkal, N., De Marco, O., et al., 2002, \textit{ApJ}, 581, L39
\bibitem[Lawlor \& MacDonald (2003)]{Lawlor03}
Lawlor, T. M., \& MacDonald, J., 2003, \textit{ApJ}, 583, 913
\bibitem[Miller Bertolami  et al. (2006)]{MillerBertolami06}
Miller Bertolami, M. M., Althaus, L. G., Serenelli, A. M., Panei, J. A., 2006, \textit{A\&A}, 449, 313
\bibitem[Nakano et al. (1996)]{Nakano96}
Nakano, S., Sakurai, Y., Hazen, M., et al., 1996, \textit{IAU Circ.}, 6322, 1
\bibitem[van Hoof et al. (2007)]{vHoof07}
van Hoof, P. A. M., Hajduk, M., Zijlstra, A. A., et al., 2007, \textit{A\&A}, 471, L9
\bibitem[van Hoof  et al. (2008)]{vHoof08}
van Hoof, P. A. M., Hajduk, M., Zijlstra, A. A., et al., 2008, \textit{ASP Conf. Ser.}, vol.~391, 155 

\end{thebibliography}
\end{document}